\documentclass{article}
\usepackage{arxiv}

\usepackage[utf8]{inputenc} 
\usepackage[T1]{fontenc}    
\usepackage{hyperref}       
\usepackage{xcolor}
\usepackage{url}
\usepackage{graphicx}

\usepackage{cite}
\usepackage{amsmath,amssymb,amsfonts,nccmath}
\usepackage{algorithmic}
\usepackage{textcomp}
\usepackage{siunitx}
\usepackage[caption=false,font=footnotesize]{subfig}

\hypersetup{
    colorlinks=true, 
    linkcolor=blue, 
    citecolor=red, 
    filecolor=black, 
    urlcolor=black   
}

\title{Employing Vector Field Techniques on the Analysis of Memristor Cellular Nonlinear Networks Cell Dynamics}
\date{September 30, 2023}

\author{ Chandan~Singh \\
    Faculty of Electrical and Computer Engineering\\
    Technische Universit\"at Dresden\\
    Dresden, Germany\\
	\texttt{chandan\_omprakash.singh@mailbox.tu-dresden.de} \\
	\And
	\href{https://orcid.org/0000-0002-2367-5567}{\includegraphics[scale=0.06]{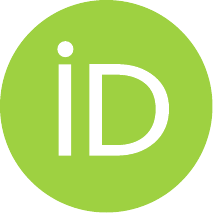}\hspace{1mm}Vasileios~Ntinas}\thanks{Presented at the 18th IEEE International Workshop on Cellular Nanoscale Networks and their Applications (\emph{CNNA'23}) and the 8th Memristor and Memristive Symposium.} \\
    Faculty of Electrical and Computer Engineering\\
    Technische Universit\"at Dresden\\
    Dresden, Germany\\
	\texttt{vasileios.ntinas@tu-dresden.de} \\
	\And
	\href{https://orcid.org/0000-0002-6200-4707}{\includegraphics[scale=0.06]{orcid.pdf}\hspace{1mm}Dimitrios~Prousalis} \\
    Faculty of Electrical and Computer Engineering\\
    Technische Universit\"at Dresden\\
    Dresden, Germany\\
	\texttt{dimitrios.prousalis@tu-dresden.de} \\
	\And
	\href{https://orcid.org/0000-0002-9560-9542}{\includegraphics[scale=0.06]{orcid.pdf}\hspace{1mm}Yongmin~Wang} \\
    Peter Gr\"unberg Institut (PGI-10)\\
    Forschungszentrum J\"ulich GmbH\\ 
    J\"ulich, Germany\\
	\texttt{yon.wang@fz-juelich.de} \\
	\And
	\href{https://orcid.org/0000-0002-1236-1300}{\includegraphics[scale=0.06]{orcid.pdf}\hspace{1mm}Ahmet~Samil~Demirkol} \\
    Faculty of Electrical and Computer Engineering\\
    Technische Universit\"at Dresden\\
    Dresden, Germany\\
	\texttt{ahmet\_samil.demirkol@tu-dresden.de} \\
	\And
	\href{https://orcid.org/0000-0002-4286-6553}{\includegraphics[scale=0.06]{orcid.pdf}\hspace{1mm}Ioannis~Messaris} \\
    Faculty of Electrical and Computer Engineering\\
    Technische Universit\"at Dresden\\
    Dresden, Germany\\
	\texttt{ioannis.messaris@tu-dresden.de} \\
	\And
	\href{https://orcid.org/0000-0001-5432-0286}{\includegraphics[scale=0.06]{orcid.pdf}\hspace{1mm}Vikas~Rana} \\
    Peter Gr\"unberg Institut (PGI-10)\\
    Forschungszentrum J\"ulich GmbH\\ 
    J\"ulich, Germany\\
	\texttt{v.rana@fz-juelich.de} \\
	\And
	\href{https://orcid.org/0000-0002-4258-2673}{\includegraphics[scale=0.06]{orcid.pdf}\hspace{1mm}Stephan~Menzel} \\
    Peter Gr\"unberg Institut (PGI-7)\\
    Forschungszentrum J\"ulich GmbH\\ 
    J\"ulich, Germany\\
	\texttt{st.menzel@fz-juelich.de} \\
	\And
	\href{https://orcid.org/0000-0003-4026-9648}{\includegraphics[scale=0.06]{orcid.pdf}\hspace{1mm}Alon~Ascoli} \\
    Faculty of Electrical and Computer Engineering\\
    Technische Universit\"at Dresden\\
    Dresden, Germany\\
	\texttt{alon.ascoli@tu-dresden.de} \\
	\And
	\href{https://orcid.org/0000-0001-7436-0103}{\includegraphics[scale=0.06]{orcid.pdf}\hspace{1mm}Ronald~Tetzlaff} \\
    Faculty of Electrical and Computer Engineering\\
    Technische Universit\"at Dresden\\
    Dresden, Germany\\
	\texttt{ronald.tetzlaff@tu-dresden.de}
}

\renewcommand{\headeright}{Chandan~Singh \it{et. al}}
\renewcommand{\undertitle}{Presented Conference Paper}
\renewcommand{\shorttitle}{IEEE CNNA 2023}

\hypersetup{
pdftitle={Employing Vector Field Techniques on the Analysis of Memristor Cellular Nonlinear Networks Cell Dynamics},
pdfsubject={cs.ET, MCNN},
pdfauthor={Chandan~Singh, Vasileios~Ntinas, Dimitrios~Prousalis, Yongmin~Wang, Ahmet~Samil~Demirkol, Ioannis~Messaris,
Vikas~Rana, Stephan~Menzel, Alon~Ascoli, Ronald~Tetzlaff},
pdfkeywords={Memristor Cellular Nonlinear Networks, Vector Field, Valence Change Mechanism Memristors},
}

\begin{document}

\maketitle

\begin{abstract}
This paper introduces an innovative graphical analysis tool for investigating the dynamics of Memristor Cellular Nonlinear Networks (M-CNNs) featuring 2\textsuperscript{nd}-order processing elements, known as M-CNN cells. In the era of specialized hardware catering to the demands of intelligent autonomous systems, the integration of memristors within Cellular Nonlinear Networks (CNNs) has emerged as a promising paradigm due to their exceptional characteristics. However, the standard Dynamic Route Map (DRM) analysis, applicable to 1\textsuperscript{st}-order systems, fails to address the intricacies of 2\textsuperscript{nd}-order M-CNN cell dynamics, as well the 2\textsuperscript{nd}-order DRM (DRM2) exhibits limitations on the graphical illustration of local dynamical properties of the M-CNN cells, e.g. state derivative's magnitude. To address this limitation, we propose a novel integration of M-CNN cell vector field into the cell's phase portrait, enhancing the analysis efficacy and enabling efficient M-CNN cell design. A comprehensive exploration of M-CNN cell dynamics is presented, showcasing the utility of the proposed graphical tool for various scenarios, including bistable and monostable behavior, and demonstrating its superior ability to reveal subtle variations in cell behavior. Through this work, we offer a refined perspective on the analysis and design of M-CNNs, paving the way for advanced applications in edge computing and specialized hardware.
\end{abstract}


\section{Introduction}
In the realm of computing, we find ourselves at the precipice of a significant paradigm shift. Traditional general-purpose computing systems are giving way to a new breed of specialized hardware, designed with utmost efficiency to tackle specific tasks. The driving force behind this transformative trend can be attributed to the heightened need for intelligent autonomous systems, specifically edge devices, which are capable of conducting sensing and processing tasks remotely. Consequently, these devices are constrained by a restricted power allocation and limited physical space.

The Cellular Nonlinear Network (CNN) is a widely recognized computer architecture that exhibits high levels of parallelism and possesses universal processing capabilities \cite{chua1998cnn}. Standard CNN implementations are frequently employed for high-speed image processing tasks, which can be efficiently integrated into arrays of photo-sensors \cite{chua2002cellular,rodriguez2018cmos}. In recent times, there has been a surge in interest surrounding memristors, which are novel electrical devices at the nanoscale level. These devices have garnered attention due to their exceptional characteristics, such as non-volatility, adaptability, low-power operation, and the ability to achieve high levels of integration \cite{tetzlaff2013memristors,siemon2015realization,ascoli2022deep}. The integration of memristor devices within CNN cells facilitated the storage of localized information within individual cells, thereby enhancing their dynamic characteristics \cite{tetzlaff2019theoretical,ascoli2019theoretical,messaris2020multi,ascoli2020theoretical}.

Aiming to harness the unlimited processing capabilities of CNNs, Chua developed the Dynamic Route Map (DRM) representation as a theoretical tool to analyze 1\textsuperscript{st}-order dynamical circuit behavior \cite{chua2018five}. By utilizing this tool, individuals can effectively design and analyze the functioning of a CNN cell as a 1\textsuperscript{st}-order dynamical system, thereby enabling the development of distinct processing operations. Moreover, a comprehensive analysis of a class of Memristor CNNs (M-CNN) has been presented in \cite{tetzlaff2019theoretical}. M-CNNs are obtained through the modification of conventional CNNs by incorporating a singular non-volatile memristor in parallel to the resistor-capacitor pair within each CNN cell. Considering the fact that the state evolution of the memristor contributes to the enhancement of cell's functionality, and taking into account the limitation of DRM-based analysis to 1\textsuperscript{st}-order dynamical systems \cite{ntinas2023design}, it becomes apparent that examining the functionalities of 2\textsuperscript{nd}-order processing elements within the aforementioned class with the DRM is no longer effective. Hence, the scope of the DRM graphical tool has been expanded to encompass nonlinear dynamical systems that possess two degrees of freedom \cite{tetzlaff2019theoretical}. This extension, referred to as 2\textsuperscript{nd}-order DRM (DRM2) \cite{tetzlaff2019theoretical}, draws inspiration from the Phase Portrait concept within the realm of nonlinear dynamics theory and illustrates the different regions within cell's phase plane on the basis of the sign of each cell state derivative. Nevertheless, the aforementioned tools do not incorporate the visualization of crucial information, such as the magnitude of each derivative that governs the local behavior of the cell.

In this present study, we aim to leverage the existing knowledge of the DRM and DRM2 analyses and tackle their limitations in order to achieve a more comprehensive expansion of these invaluable tools. In specific, we propose the integration of M-CNN cell's vector field within the graphical representation of the cell's Phase Portraits, thereby enhancing the efficacy of the design tool for the intricately dynamic M-CNN cells. Furthermore, we employ an accurate physics-based memristor model in order to face challenges associated with designing realistic M-CNN cells. These challenges include the logarithmic scale of memristor's state variable and significant magnitude scaling issues, which pose considerable difficulties in visualizing the vector field. The graphical analysis tool proposed here allows for a direct understanding of M-CNN cell's behavior and facilitates its efficient design.

\section{M-CNN Cell Dynamics}

\begin{figure}[!t]
\centering
\includegraphics[width=0.85\linewidth]{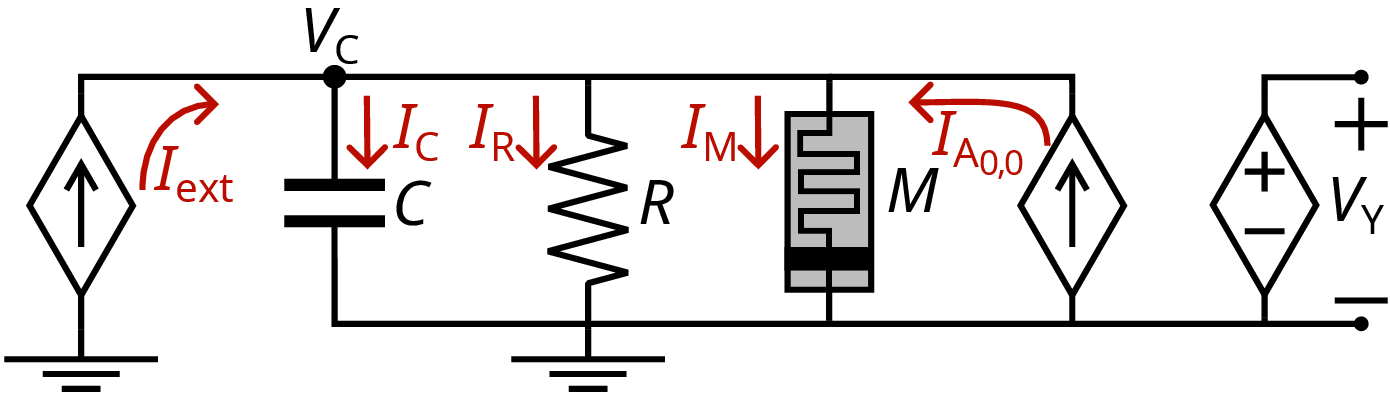}
\caption{(a) M-CNN Cell's equivalent circuit.}
\label{fig:my_chua_cell}
\end{figure}

Chua and Yang introduced the CNN architecture in their influential publication \cite{chua1988cellular}. This framework involves a grid of interconnected cells that possess the capability to carry out computational tasks with high efficiency. Each CNN cell $C_{i,j}$ consists of several components, including a resistor $R$, a capacitor $C$, and a self-feedback-dependent current source $I_{A_{0,0}}$, all connected in parallel, and an output voltage source $V_{Y}$ (Fig.~\ref{fig:my_chua_cell}). The $R-C$ component is employed to form the state variable of each cell by charging and discharging the capacitor voltage $V_{C}$, while the source $V_{Y}$ serves as the output signal of each cell, typically representing a nonlinear mapping of the state variable to a bounded output signal. Moreover, the self-feedback $I_{A_{0,0}}$ plays an important role on the computational operations of the CNN cell by regulating the stability of the cell.

The CNN cell functionality is completed by the extracell contributions, which consist of the weighted signals from the neighboring cells' inputs $V_{U|i,j}(t)$ and outputs $V_{Y|i,j}(t)$ and a fixed bias $I_{{\rm z|}i,j}$, which are calculated as

\begin{equation}
\begin{split}
    I_{ext|i,j}(t) = &\sum_{\substack{k=i-1,l=j-1\\(k,l)\neq(i,j)}}^{i+1,j+1} A_{k-i,l-j}V_{Y|k,l}(t) + \nonumber\\
    &\sum_{k=i-1,l=j-1}^{i+1,j+1} B_{k-i,l-j}V_{U|k,l}(t) + I_{\rm z}.
    \label{eq:offset_current}
\end{split}
\end{equation}

The coefficients $B_{k-i,l-j}$ and $A_{k-i,l-j}$ are the coupling weights of $C_{i,j}$ with the input and output of $C_{k,l}$, respectively, where $k\in\{i-1,i,i+1\}$ and $l\in\{j-1,j,j+1\}$ for the standard CNN structure. Moreover, a factor $z$ defines the independent current bias, such as $I_{\rm z}=z$. For a space-invariant CNN, the complete set of weights $A_{k,l}$, $B_{k,l}$, and $z$ for $k,l\in\{-1,0,1\}$ defines the CNN template, and its proper selection leads to different computational tasks.

An enhancement of conventional CNNs capabilities is achieved by integrating a distinct non-volatile memristor $M$ in parallel to the $R$-$C$ pair within each CNN cell, forming the M-CNN architecture. Considering a realistic memristive device, like the Valence Change Mechanism (VCM) devices modeled by the Juelich-Aachen Resistive Switching Tool (JART) \cite{bengel2020variability}, the voltage- and state-dependent resistance of the cell's memristor $R_{M}$ should be incorporated into cell's dynamics, constituting the following 2\textsuperscript{nd}-order dynamical system


\begin{equation}
\dot{V_C}=\frac{dV_C}{dt} = \frac{I_{ext|i,j}(t)+I_{A_{0,0}}}{C}-\frac{V_C}{RC} - \frac{V_C}{R_{M}\left(V_C,N_d\right)C} \label{eq:dvdt}
\end{equation}

\begin{equation}
\dot{N_d}=\frac{dN_{d}}{dt} = - \frac{I_{ion}\left(V_C,N_d\right)}{{z_{\rm V_O}eAl_{d}}}\label{eq:dNddt}
\end{equation}

\noindent where $N_d$ is the state variable of the JART VCM memristor model, representing the concentration of oxygen vacancies in the vicinity of the interface between the switching layer and the active electrode of the device, named as the \emph{disc region}. Moreover, $l_{d}$ denotes the length of the disc region, $z_{\rm V_{O}}$ stands for the oxygen vacancy charge number, $e$ represents the elementary charge and $A=\pi{}r_{d}^2$ defines the area of the switching filament. The ionic current $I_{ion}$ represents the current of the moving oxygen vacancies during memristor's switching and it depends on $N_d$ and the applied voltage, as elaborated in \cite{ntinas2022toward}. For the calculation of $R_{M}$, the explicit memristor current equation from the simplified JART VCM model \cite{ntinas2023simplified} is used since it offers faster and more stable memristor current calculation.

\section{M-CNN Cell Dynamics Analysis}
Aiming to delve deeper into the cell dynamics, enriched by the incorporation of a memristor device, the context of this work is focused on the isolated M-CNN cell, where $I_{ext|i,j}(t)=0$, so all the template weights are zero, apart from the cell's self-feedback weight $A_{0,0}$.

Let us consider a typical M-CNN cell configuration, where $I_{A_{0,0}}=A_{0,0}V_{Y}/R$ with $A_{0,0}=2$ and 

\begin{equation}
    V_{Y}=F_{Y}(V_C)=\frac{1}{2}\left(\left|V_C+1\right|-\left|V_C-1\right|\right).
\end{equation}

\begin{figure}[!t]
\centering
\subfloat[]{
\includegraphics[width=0.47\linewidth]{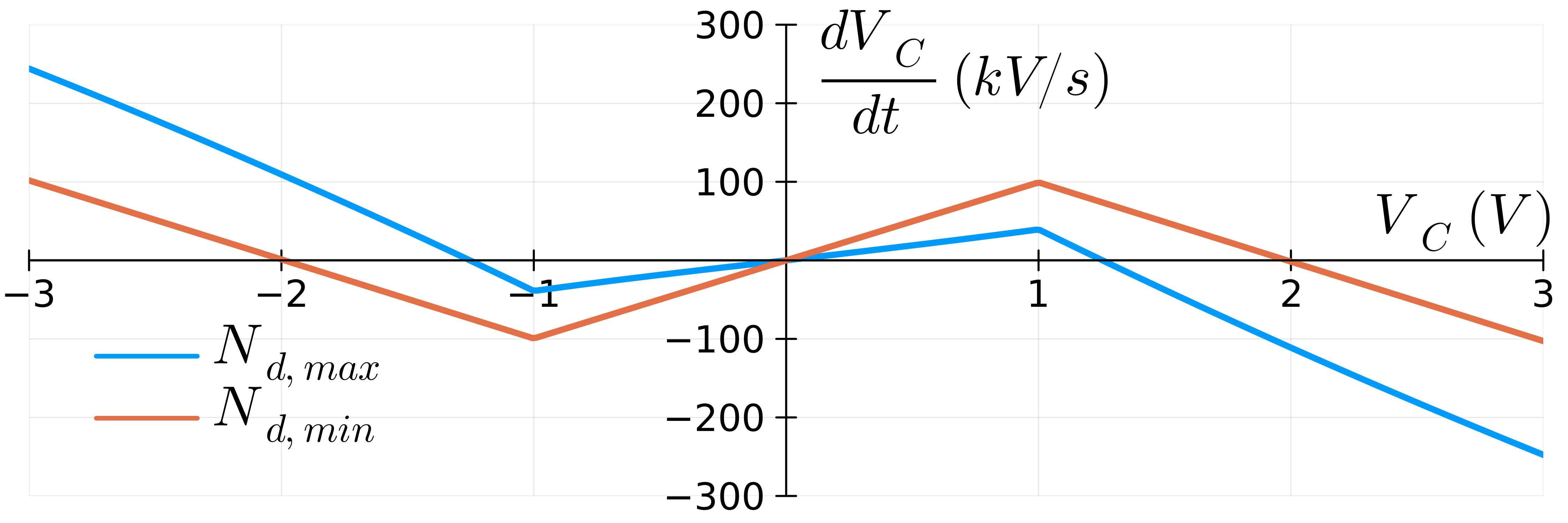}}
\subfloat[]{
\includegraphics[width=0.47\linewidth]{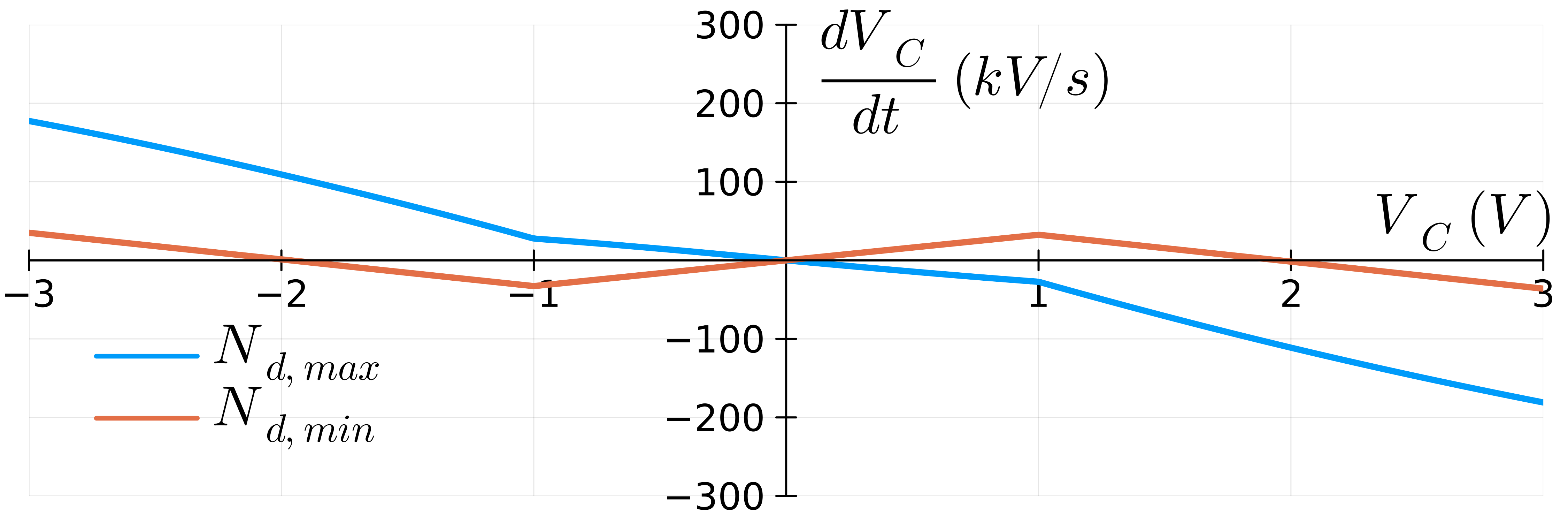}}
\caption{Illustration of M-CNN Cell's SDRs for $N_d=N_{d,min}$ and $N_d=N_{d,max}$ with $R$ selected as (a) $R=\qty{1}{\kilo\Omega}<R_{M,min}<R_{M,max}$ and, (b) $R_{M,min}<R=\qty{3}{\kilo\Omega}<R_{M,max}$.}
\label{fig:DRM}
\end{figure}

If $R$ is properly selected such as $R<R_{M,min}<R_{M,max}$\footnote{When $N_d$ is at its lowest (highest) value $N_{d,min}$ ($N_{d,max}$) we observe $R_{M,max}$ ($R_{M,min}$).}, the M-CNN cell exhibits bistable behavior regardless memristor's state. In that case, both the State Dynamic Route (SDR) for the lowest and the highest $N_d$ values, shown in Fig.~\ref{fig:DRM}(a), consist of a three-piece piecewise linear function with the middle piece having positive slope. Employing, first, conventional CNN analysis techniques, we investigate cell's equilibria, where $\dot{V_C}=0V/s$, for theoretical non-dynamical memristors, i.e. $\dot{N_d}=0$ always. So, each of those SDRs shows one unstable equilibrium at $V_C=0V$ and two stable equilibria, one at negative and one at positive $V_C$ values.

On the other hand, if $R$ is selected such as $R_{M,min}<R<R_{M,max}$, the SDR for $N_d=N_{d,max}$ changes shape such that the middle piece admits negative slope, as shown in Fig.~\ref{fig:DRM}(b). As a result, the M-CNN cell for $N_d=N_{d,max}$ has only one equilibrium point at $V_C=0V$ and it is a stable one.

Memristors are dynamical elements and the above analysis method is limited to the cases that memristor's dynamics are much faster than capacitor's ones, such that memristor is abruptly switching between $N_{d,min}$ and $N_{d,max}$. In that case, the whole M-CNN cell's behavior can be illustrated by the transition from one SDR to the other, when memristor switches. For M-CNN cells with fast capacitors or/and memristor devices with gradual switching properties, an analysis technique that incorporates the whole range $N_{d,min}\leq N_{d} \leq N_{d,max}$ should be employed.

\begin{figure}[!t]
\centering
\subfloat[]{
\includegraphics[width=0.49\linewidth]{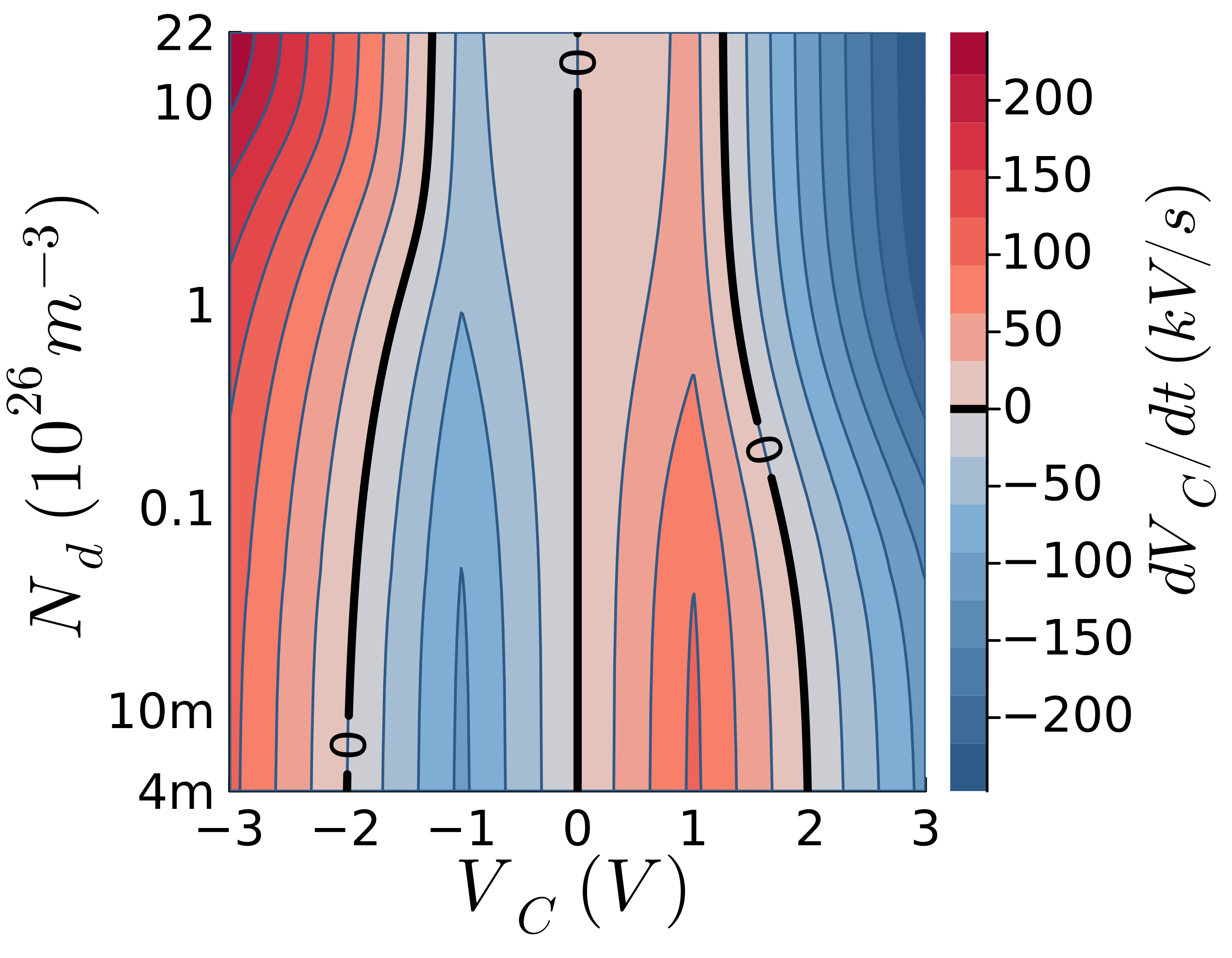}}
\subfloat[]{
\includegraphics[width=0.49\linewidth]{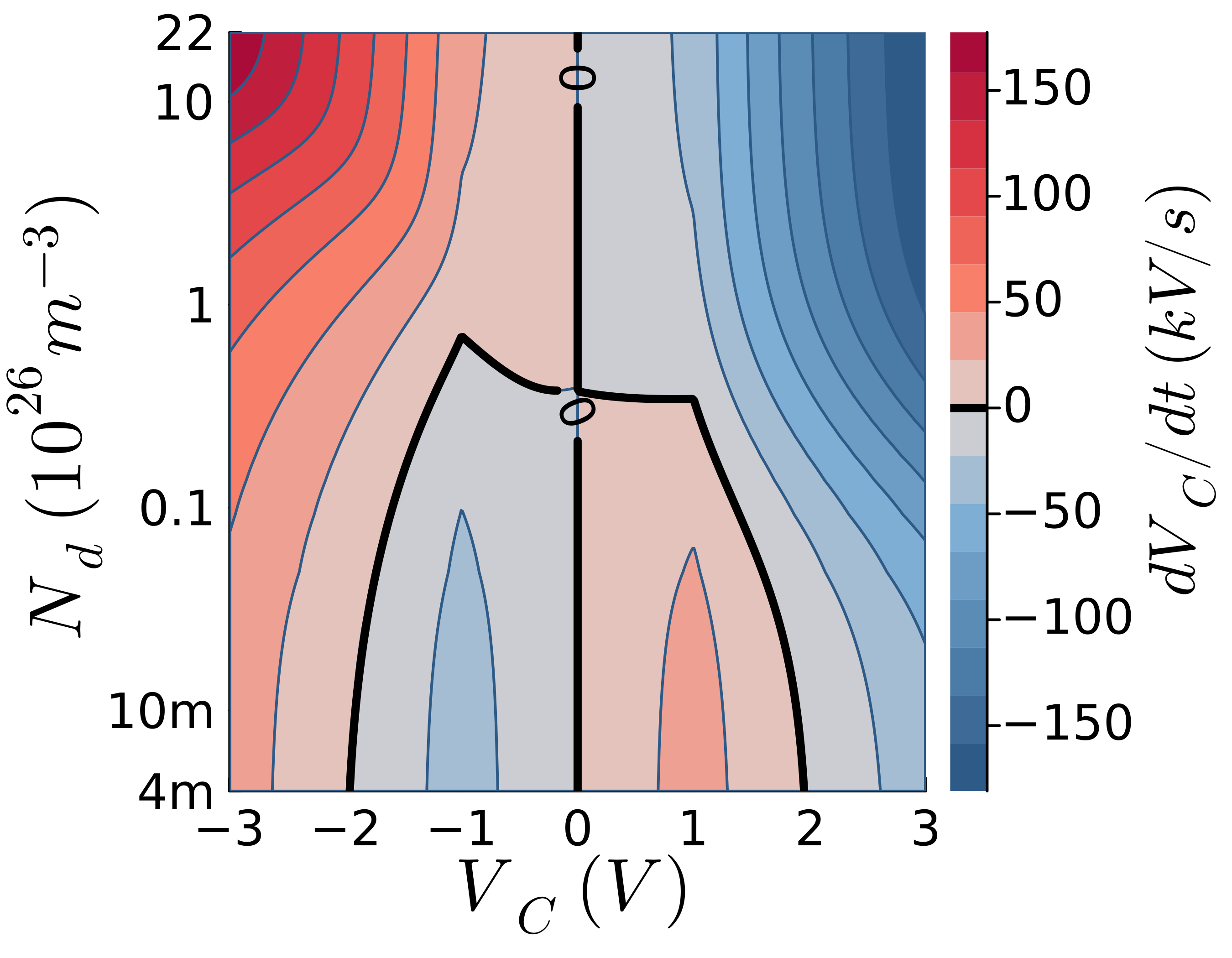}}
\\ 
\subfloat[]{
\includegraphics[width=0.49\linewidth]{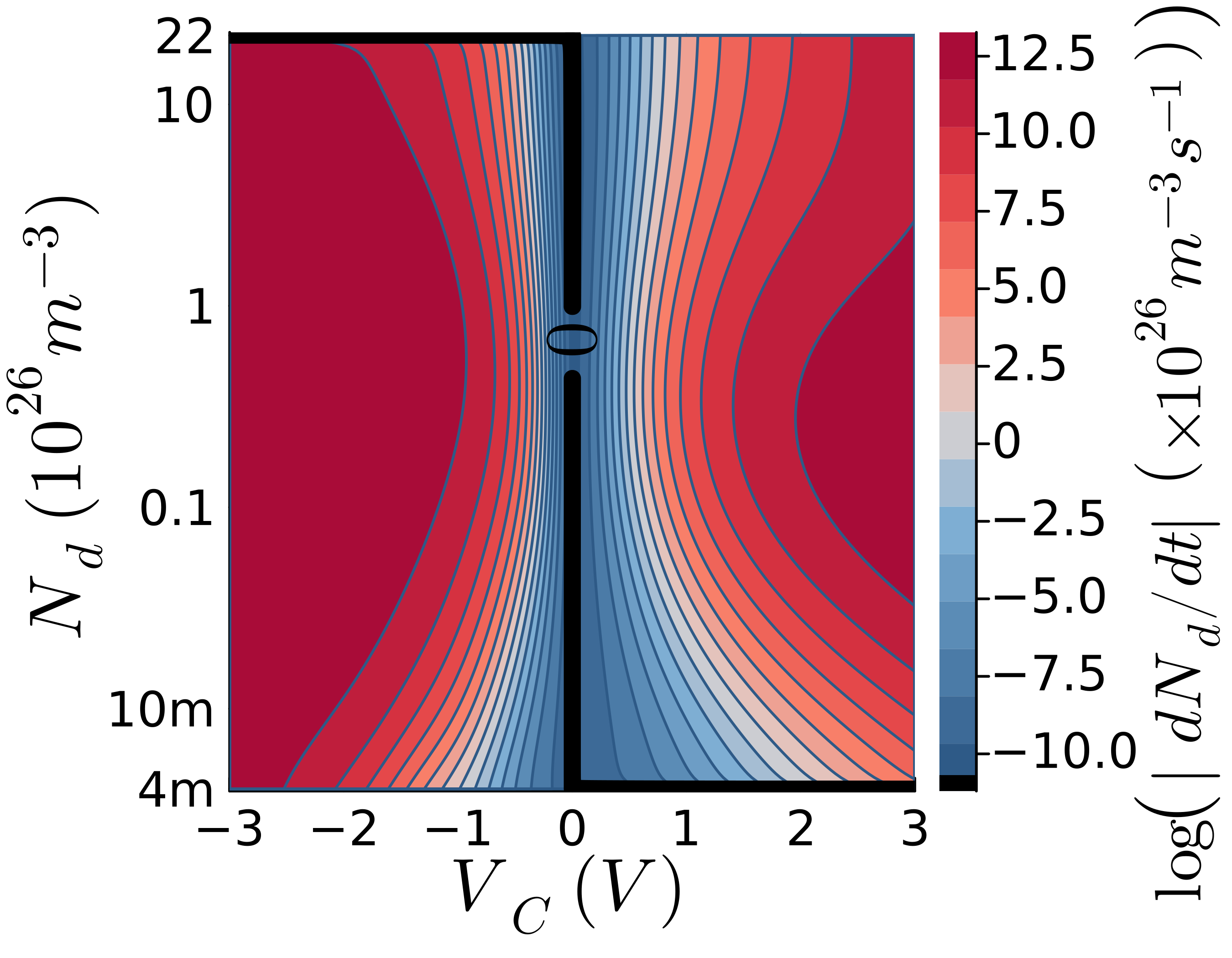}}
\caption{Capacitor voltage derivative for (a) $R=\qty{1}{\kilo\Omega}<R_{M,min}<R_{M,max}$, and (b) $R_{M,min}<R=\qty{3}{\kilo\Omega}<R_{M,max}$, and (c) memristor state derivative within M-CNN cell's phase plane.}
\label{fig:2d_Dynamics}
\end{figure}

In that case, where M-CNN cell cannot be simplified to a 1\textsuperscript{st}-order system, the 2\textsuperscript{nd}-order system \eqref{eq:dvdt}-\eqref{eq:dNddt} should be visualized in its complete phase plane, where $N_{d}\in[N_{d,min}, N_{d,max}]$ and $V_{C}\in[V_{C,min}, V_{C,max}]$. Since there is no strict limitation on $V_C$ value, we select here $V_{C}\in[-3V, 3V]$. Fig.~\ref{fig:2d_Dynamics} illustrates separately the capacitor dynamics (Fig.~\ref{fig:2d_Dynamics}(a)-(b)) and the memristor dynamics (Fig.~\ref{fig:2d_Dynamics}(c)) in the 2d phase plane. The dynamics of cell's capacitor and memristor are interdependent, so, their separate illustration does not allow for the analysis of M-CNN cell's behavior.

For that reason, the DRM2 tool \cite{tetzlaff2019theoretical} combines them into a single illustration by using the sign of each derivative, leading to four regions with all the combinations of positive and negative capacitor's voltage derivative and memristor's state derivative. Unfortunately, by only using the sign, the information on the magnitude of each derivative is missing and, as a result, the DRM2 tool does not facilitate the direct visualization of M-CNN cell's dynamical behavior, requiring for the addition of trajectory lines on the same plot, which are generated by simulating the cell for various initial conditions.

In this paper, we extend this graphical analysis tool by incorporating the illustration of cell's vector field into its phase plane. The vector field of a 2\textsuperscript{nd}-order system refers to the mapping that describes how the derivatives of the state variables depend on the current state of the system. This current state refers to each $\{V_C,N_d\}$ pair in the phase plane, while the local derivative values define the magnitude and the angle of an arrow (vector) that shows the direction along which the cell state will evolve.

\begin{figure*}[!t]
\centering
\subfloat[]{
\includegraphics[width=0.32\linewidth]{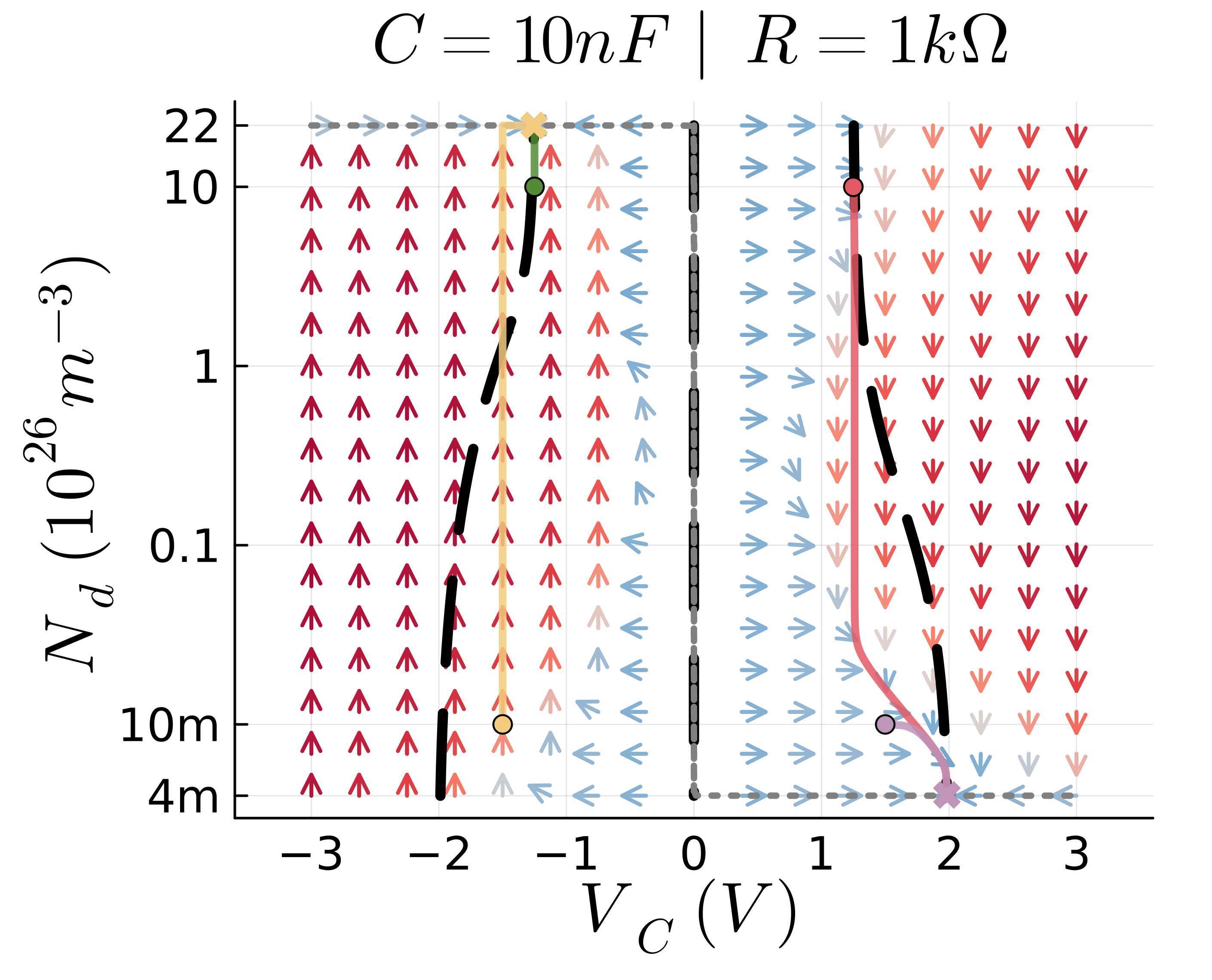}}
\subfloat[]{
\includegraphics[width=0.32\linewidth]{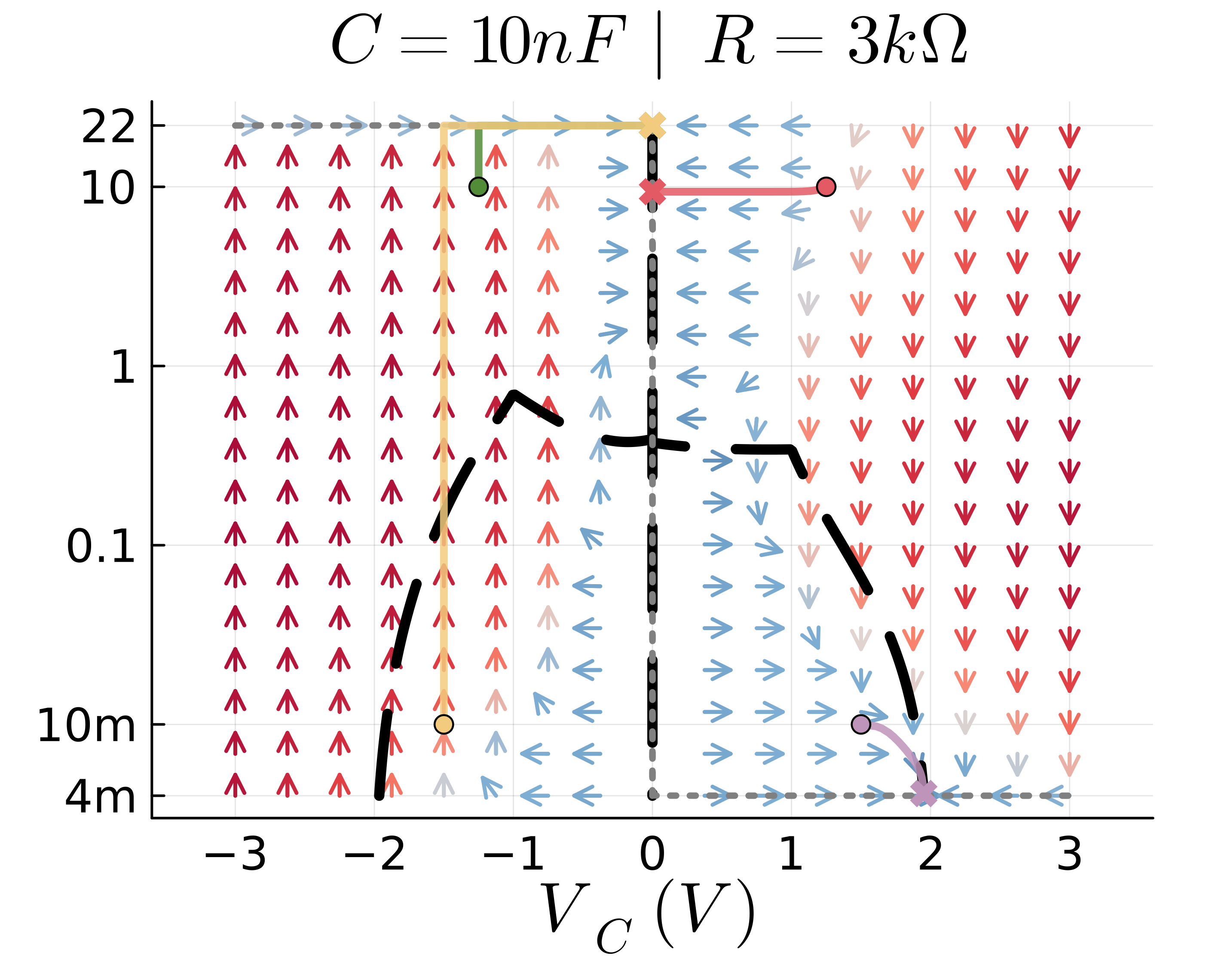}}
\subfloat[]{
\includegraphics[width=0.32\linewidth]{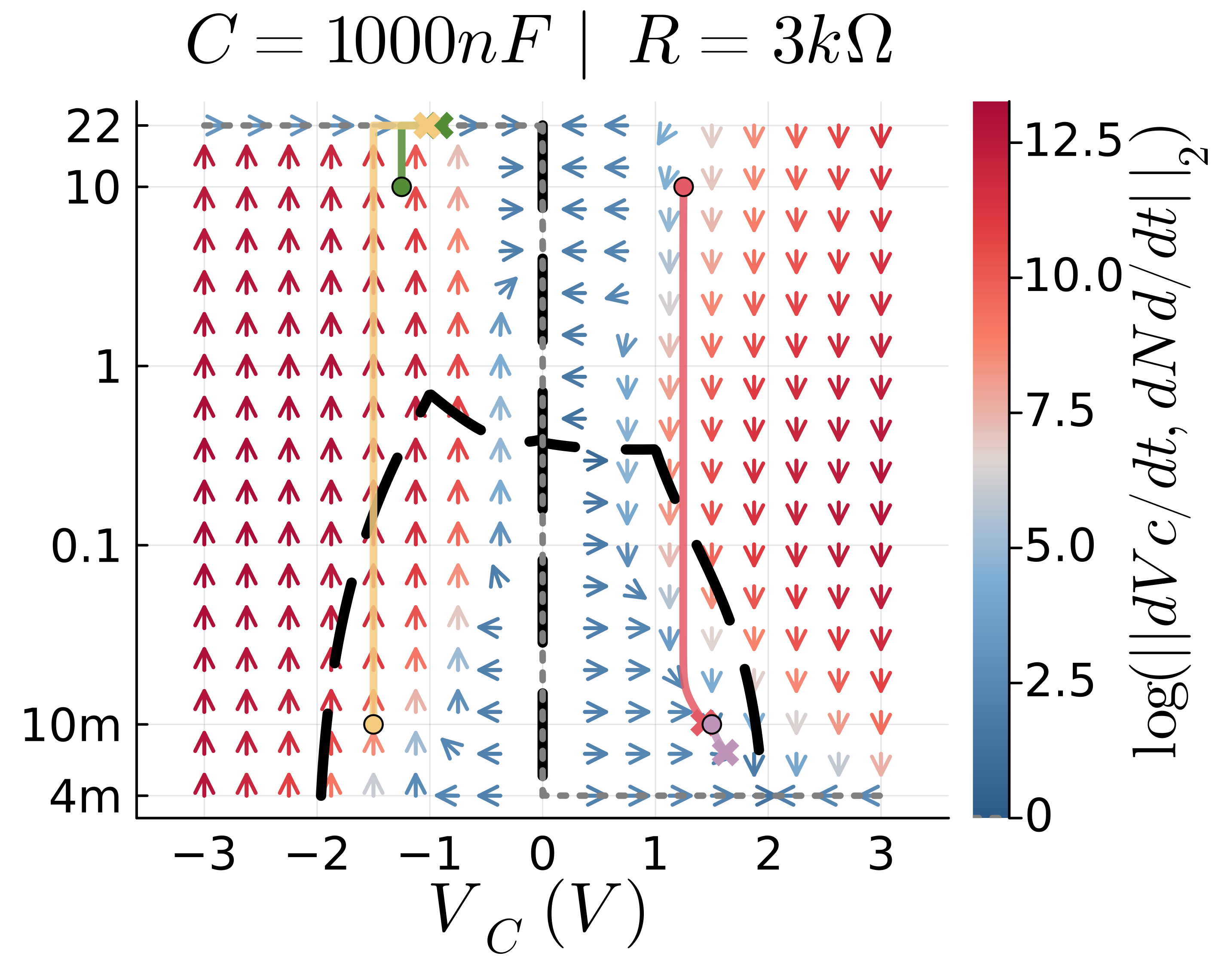}}
\caption{Enhanced graphical illustration of M-CNN cell's dynamics including the cell's vector field. (a) Fully bistable M-CNN cell design. (b)-(c) Hybrid mono- and bistable M-CNN cell design for different $C$ values. In all plots, the $V_C$ and $N_d$ nullclines are illustrated in blue dashed and gray dotted lines, respectively, while the solid lines are cell trajectories starting from the circle marker at $\{V_{C,0}, N_{d,0}\}$ and end at the cross marker at $\{V_{C,t}, N_{d,t}\}$ after a simulation with duration $t=\qty{1}{\m\s}$. 
}
\label{fig:arrow}
\end{figure*}

Deriving the vector field for the M-CNN cell dynamics constitutes a tricky task since the two state equations exhibit big scaling differences, as well as memristor dynamics individually shows differences of many orders of magnitude for different $V_C$ values. Based on that, we propose the normalization of arrow's magnitude, since information cannot be efficiently coded on its exact value. In our approach, all the vectors' lengths are the same, lying always on an ellipse with fixed vertical and horizontal radii, while the arrow's angle is calculated via the two state derivatives, such that

\begin{equation}
    \theta\left(V_C, N_d\right)=tan^{-1}\left(\frac{\dot{N_d}\left(V_C, N_d\right)}{\dot{V_C}\left(V_C, N_d\right)}\right)
\end{equation}

Moreover, in order to incorporate the missing information of vector's magnitude into the arrow, we encode it to the coloring of the arrow. Such that, the arrow color is selected based on the 2-norm, also known as Euclidean norm, of the vector at each phase plane point.

In addition, moving from the 1\textsuperscript{st}-order DRM where the zero crossings constitute equilibrium points, the proposed 2\textsuperscript{nd}-order illustration shows also the corresponding points, known as nullclines, for $\dot{V_C}$ by dashed black line and $\dot{N_d}$ by dotted gray line. The equilibria for the 2\textsuperscript{nd}-order cell appear at the crossings of the $V_{C}$ and $N_d$ nullclines. The points along the locus $dVc/dt=0$, which correspond to stable equilibria for the CNN cell (as well as for the M-CNN cell, in those scenarios where its memristor operates as an adynamic resistive element \cite{tetzlaff2019theoretical}), attract locally phase-plane solution trajectories for the \textsuperscript{nd}-order M-CNN cell. 
The plots in Fig.~\ref{fig:arrow} illustrate the application of the proposed dynamical analysis graphical tool for the investigation of M-CNN cell's behavior. The colored solid lines are the trajectories of the cell starting from a selected initial $\{V_{C,0}, N_{d,0}\}$ point and simulated for certain time ($t=\qty{1}{\m\s}$).

Fig.~\ref{fig:arrow}(a) illustrates the proposed graphical analysis tool for a M-CNN cell with $R<R_{M,min}<R_{M,max}$, where the cell's dynamics exhibit always bistable behavior, which is evident from the shape of the nullclines. On the other hand, Fig.~\ref{fig:arrow}(b) demonstrates the case where $R_{M,min}<R<R_{M,max}$, and mono- and bistability coexist. In that case, the arrows directly illustrate the direction that cell states will follow, making evident the transition path from bistability to monostability of capacitor voltage's dynamics. One more example is shown in Fig.~\ref{fig:arrow}(c), where the cell's design is the same, apart from the capacitance value of cell's capacitor. That change affects the relationship between capacitor's and memristor's switching speed, leading to a direct change to some arrows' angle. As it is evident when comparing Fig.~\ref{fig:arrow}(b) with Fig.~\ref{fig:arrow}(c), the trajectories of some initial conditions exhibit different behaviors, e.g. the red lines ($\{V_{C,0}=1.25V, N_{d,0}=10\times{}10^{26}m^{-3}\}$). Those different behaviors are directly illustrated by our proposed method, however, they could not be shown in a DRM, but also the DRM2 tool is not able to naturally illustrate them, since it is limited to the sign of the derivatives only.

\section{Conclusions}
In the present manuscript, we propose a novel graphical representation for the purpose of analyzing the 2\textsuperscript{nd}-order M-CNN cell. Drawing inspiration from the widely recognized DRM and DRM2 graphical illustration tools, a novel and sophisticated tool is postulated, wherein the system's vector field are seamlessly integrated into the complete phase plane. With the utmost intention of accentuating the merits inherent in the proposed methodology, we hereby present a comprehensive exposition of the isolated M-CNN cell's dynamics across a multitude of scenarios, wherein the advantages of the vector field representation are unequivocally discernible.

\section{Acknowledgments}
This work was supported in part by the Deutsche Forschungsgemeinschaft within the project SPP2622 Mem2CNN, in part by the Federal Ministry of Education and Research (BMBF, Germany) in the project NEUROTEC II (project numbers 16ME0398K and 16ME0399), and in part by NeuroSys A with the Ref.No: 03ZU1106AA.

\bibliographystyle{IEEEtran}
\bibliography{biblio}

\end{document}